\documentclass[12pt]{article}
\def\cdate{{June 5, 2025}}
\usepackage{amsfonts}
\usepackage{amsmath}
\usepackage{latexsym}
\usepackage{amssymb}
\usepackage{txfonts}
\usepackage{bm}
\usepackage[american]{babel}
\usepackage{graphicx}
\usepackage{framed}
\usepackage{xcolor}
\definecolor{mygray}{gray}{0.95} 
\definecolor{mydarkgray}{gray}{0.70} 
\colorlet{shadecolor}{mygray}
\makeatletter
\def\timenow{%
\@tempcnta=\time \divide\@tempcnta by 60 \number\@tempcnta:\multiply
\@tempcnta by 60 \@tempcntb=\time \advance\@tempcntb by -\@tempcnta
\ifnum\@tempcntb <10 0\number\@tempcntb\else\number\@tempcntb\fi}
\newcounter{outputpage}
\renewcommand{\@oddhead}
{\stepcounter{outputpage}\hfill\hfill\theoutputpage}
\renewcommand{\@evenhead}
{\stepcounter{outputpage}\hfill\hfill\theoutputpage}
\renewcommand{\@oddfoot}
{\vbox{
\hrule
\vspace{3pt}
\hfil
{\scriptsize\textit{
\hfill\hfill\jobname.tex; \today; \timenow; p. \theoutputpage}}
\hfil
}}
\renewcommand{\@evenfoot}
{\vbox{
\hrule
\vspace{3pt}
\hfil
{\scriptsize\textit{
\hfill\hfill\jobname.tex; \today; \timenow; p. \theoutputpage
}}
\hfil
}}
\makeatother

\def\nmt{{
\null
\vspace{-4cm}
\par
\hspace*{50truemm}{\hrulefill}
\par
\vskip-4truemm
\par
\hspace*{50truemm}{\hrulefill}
\par\vskip5mm
\par
\hspace*{50truemm}{{\large\sc
New Mexico Tech {\rm
(\cdate)
}}}\vskip4mm
\par
\hspace*{50truemm}{\hrulefill}
\par
\vskip-4truemm
\par
\hspace*{50truemm}{\hrulefill}
\par}}

\def\RR{\mathbb{R}}

\def\ZZ{\mathbb{Z}}

\def\Det{\mathrm{Det\,}}

\def\dim{\mathrm{dim\,}}

\def\Imag{\mathrm{Im\,}}

\def\Ker{\mathrm{Ker\,}}

\def\Res{\mathrm{Res\,}}
\def\Real{\mathrm{Re\,}}

\def\Tr{\mathrm{Tr\,}}
\def\vol{\mathrm{vol\,}}
\def\nn{{\nonumber}}
\def\be{\begin{equation}}
\def\ee{\end{equation}}
\def\bea{\begin{eqnarray}}
\def\eea{\end{eqnarray}}
\def\bed{\begin{definition}{\ }}
\def\eed{\end{definition}}
\def\ed{\end{document}}
\def\bp{\begin{proposition}}
\def\ep{\end{proposition}}
\def\bc{\begin{center}}
\def\ec{\end{center}}
\def\bi{\begin{itemize}}
\def\ei{\end{itemize}}
\def\benum{\begin{enumerate}}
\def\eenum{\end{enumerate}}
\def\bmp{\begin{minipage}}
\def\emp{\end{minipage}}

\newtheorem{proposition}{Proposition}

\newtheorem{definition}{Definition}
\begin{document}

\begin{titlepage}
\thispagestyle{empty}
\nmt

\bigskip
\bigskip
\bigskip
\bigskip
\centerline{\LARGE\bf Heat Kernel on Warped Products}
\bigskip
\bigskip
\centerline{\Large\bf Ivan G. Avramidi}
\bigskip
\centerline{\it Department of Mathematics}
\centerline{\it New Mexico Institute of Mining and Technology}
\centerline{\it Socorro, NM 87801, USA}
\centerline{\it E-mail: ivan.avramidi@nmt.edu}
\bigskip
\medskip

\begin{abstract}

We study the spectral properties of the scalar
Laplacian on a $n$-dimen\-sional warped product manifold
$M=\Sigma\times_f N$ with a $(n-1)$-dimensional
compact manifold $N$  without boundary,
a one dimensional manifold
$\Sigma$ without boundary and
a warping function $f\in C^\infty(\Sigma)$.
We consider two cases:
$\Sigma=S^1$ when the manifold $M$ is compact,
and $\Sigma=\RR$ when the manifold $M$ is non-compact.
In the latter case we assume that the warping function
$f$ is such that the manifold $M$ has two cusps
with a finite volume.
In particular, we study the case of the
warping function $f(y)=[\cosh(y/b)]^{-2\nu/(n-1)}$
in detail,
where $y\in\RR$
and $b$ and $\nu$ are some positive parameters.
We study the properties of  the spectrum of the Laplacian
in detail and show that it has both the discrete and
the continuous spectrum.
We compute the resolvent,
the eigenvalues, the scattering matrix,
the heat kernel and the regularized heat trace.
We compute the asymptotics of the regularized heat trace
of the Laplacian on the warped manifold $M$
and show that some of its coefficients are global
in nature expressed in terms of the zeta function on the
manifold $N$.

\end{abstract}

\end{titlepage}


\section{Introduction}
\setcounter{equation}{0}

Spectral theory of self-adjoint elliptic partial differential operators
of Laplace type on Riemannian manifolds
plays an important role in mathematical physics, geometric
analysis, differential geometry and quantum field theory.
The spectrum of such differential operators is most
conveniently studied by analyzing the resolvent and the spectral
functions such as the heat trace and the zeta function
\cite{gilkey95,avramidi00}.
The central problem of spectral geometry is the question: ``To what extent
does the spectrum of the Laplacian determines the geometry and the topology
of the manifold?''.

In the compact setting, for manifolds without boundary,
the spectral theory of Laplacian
is well understood.
Many problem in mathematical physics and differential geometry
lead to non-compact manifolds, see, e.g.
\cite{borthwick12}. Spectral theory of differential operators
on noncompact manifolds is more complicated;
in many respects the infinity
acts as a singularity in the compact setting
(see \cite{cheeger79,mueller83,mueller87,
bruening91,berline04,donelly79,
melrose93,albin05,vaillant01} and the references therein).

The purpose of this paper is to study the spectrum of the scalar
Laplacian, in particular, its heat kernel,
on some noncompact warped product manifolds $M=\RR\times_f N$,
where $N$ is a compact manifold without boundary
and $f\in C^\infty(\RR)$ is a smooth function decreasing at infinity
as $|y|\to\infty$ like $\exp(-c|y|)$ sufficiently fast; such manifolds have two
cusp-like ends as $y\to -\infty$ and $y\to +\infty$ and a finite volume.
Examples of such manifolds include
locally symmetric spaces of finite volume, spaces with cone-like singularities,
manifolds of finite volume with finitely many cusps etc.
\cite{mueller83,mueller87,borthwick12,vaillant01}.
In particular, if $G/K$ is a non-compact locally
symmetric space of rank one
 (with the isometry group $G$ and the isotropy group $K$)
and $\Gamma$ is a discrete subgroup of the isometry group $G$, then
the manifold $M=\Gamma\backslash G/K$ will be of this type,
that is, of finite volume with finitely
many cusps. The simplest example is the
manifold
$M = \Gamma\backslash H^2=\Gamma\backslash
SL(2,R)/SO(2)$ where $\Gamma$ is the discrete
subgroup of the group $SL(2,Z)$.
It would be interesting to apply our analysis to the study of
more general manifolds
with finitely many cusps $M_j=[a_j,\infty)\times_{f_j}N_j$,
$j=1,\dots, m$, with $a_j>0$, some smooth
warping functions $f_j$
and some compact $(n-1)$-dimensional manifolds $N_j$.

This paper is organized as follows.
In Sec. 2 we describe the geometry of the warped product manifolds
and consider specific examples of the warping function.
In Sec. 3 we introduce the heat kernel on the warped
product manifold and use the separation
of variables to construct the heat kernel
on the manifold $M$ in terms of the heat kernel of the manifold $N$.
In Sec. 4 we describe the heat trace on compact
manifolds and the related spectral functions, like the
zeta function.
In Sec. 5 we describe the asymptotic expansion of the
heat kernel diagonal of a one-dimensional Schr\"odinger
operator.
In Sec. 6
 we study the heat kernel of the relevant
Schr\"odinger operators $D_k$ with specific
confining potentials and discrete spectrum.
In Sec. 7 we study
the resolvent and the heat kernel
of the relevant operator $D_0$
with a non-confining potential and study
their properties.
We show that it has a finite number of simple discrete
eigenvalues and a continuous spectrum. We also
compute the asymptotic expansion of the regularized heat
trace of the operator $D_0$.
We consider a specific
example of a noncompact manifold of finite volume with two
cusps with a specific one-parameter family of warping functions.
We explicitly compute the resolvent,
the eigenvalues, the scattering matrix, and the heat kernel
and the heat trace.
In Sec. 8 we compute the asymptotics of the heat trace
of the Laplacian on the warped manifold $M$.


\section{ Warped Product Manifolds}
\setcounter{equation}{0}

Let $(N,h)$ be a compact $(n-1)$-dimensional
orientable Riemannian manifold without boundary
with a metric $h$.
We will find it useful to
introduce the parameter
\be
\alpha=\frac{n-1}{2},
\ee
so that $n=2\alpha+1$.
We use Latin indices to denote the local coordinates
$\hat x^i$, $i=1,\dots,n-1$, on the manifold $N$;
then the metric is given by
\be
dl^2=h_{ij}(\hat x)d\hat x^id\hat x^j.
\ee
The Riemannian volume element is defined as usual by
\bea
d\vol_N = d\hat x\,|h|^{1/2},
\eea
where
$d\hat x=d\hat x^1\wedge\cdots\wedge d\hat x^{n-1}$
and $|h|=\det h_{ij}$.
The simplest case is when the manifold $N$ is the torus
$T^{n-1}$ with zero curvature and the volume
\be
\vol(N)=(2\pi)^{2\alpha} a_1\dots a_{n-1},
\ee
where $a_i$ are the radii of the circles $S^1$.
We will also consider the case when
the manifold $N$ is just the sphere
$S^{n-1}$ of radius $a$, with the curvature
\be
F^{ij}{}_{km}
=\frac{1}{a^2}\left(\delta^i{}_k \delta^j{}_{m}
-\delta^i{}_m \delta^j{}_{k}
\right)
\ee
and the volume
\be
\vol(N)=a^{2\alpha}
\frac{2\pi^{(2\alpha+1)/2}}{\Gamma\left(\alpha+\frac{1}{2}\right)}.
\ee

Let $\Sigma$ be a one-dimensional manifold without boundary
with a local coordinate $y$.
We consider two cases:
a circle $\Sigma=S^1$ of radius $a$,
that is, $0\le y\le 2\pi a$,
and the real line $\Sigma=\RR$.
Let $f\in C^\infty(\Sigma)$ be a positive smooth function on
$\Sigma$ parameterized by
\be
f(y)=\exp\{-\omega(y)\},
\ee
A warped product manifold
\be
M=\Sigma\times_f N,
\ee
is a $n$-dimensional Riemannian manifold
$(M,g)$
with the metric
\be
ds^2=dy^2+\exp[-2\omega(y)]dl^2.
\ee
We denote the local coordinates on $M$ by $(x^\mu)=(y,\hat x^i)$,
with Greek indices running over $0,1,\dots, n-1$, and $x^0=y$.


We consider some non-compact examples with
different warping function, most importantly its behavior as $y\to \infty$.
Of course, if the warping function is constant $f(y)=c$
then we just have the cylinder $M=\RR\times N$.
If the warping function approaches a non-zero constant
at infinity $f(y)\sim c$ then we call it
a {\it cylindrical end}.
If the warping function goes to infinity at infinity
like
$
f(y)\sim\cosh(y/a)
$
then we call it a {\it funnel}.
If the warping function goes to infinity like
$f(y)\sim \sinh (y/a)$ then we call a {\it hyperbolic
cylinder}.
If the warping function goes to zero at infinity
like
$f(y)\sim [\cosh (y/a)]^{-1}\sim \exp(-|y|/a)$,
then we call it a manifold with {\it cusps}.

The determinant of the metric is
\be
|g|=\det g_{\mu\nu}= e^{-4\alpha\omega}|h|.
\ee
Therefore, the Riemannian volume element of the manifold $M$ is
\bea
d\vol_M = dx\,|g|^{1/2}
=\exp[-2\alpha\omega(y)]dy\;d\vol_N,
\eea
where
$dx=dx^0\wedge d x^1\wedge\cdots\wedge d x^{n-1}$,
and
the volume of the manifold $M$ is
\be
\vol(M)=\beta\vol(N),
\ee
where
\be
\beta
=\int\limits_{\Sigma} dy \exp[-2\alpha\omega(y)].
\label{28iga}
\ee
In the noncompact case, if
the function $\omega(y)$ increases at
infinity  $|y|\to\infty$
sufficiently fast,
\be
\omega(y)>\gamma\log |y|,
\ee
with $\gamma>1/(2\alpha)$, then
the volume of the manifold $M$
is finite.



The non-zero components of the curvature of the manifold $M$
are
\bea
R{}^{0k}{}_{0i} &=&
-\delta^{k}{}_i (\omega'^2-\omega''),
\\
R{}^{ij}{}_{km} &=& e^{2\omega}F^{ij}{}_{km}
-\left(\delta^i{}_k \delta^j{}_{m}-\delta^i{}_m \delta^j_{k}
\right)\omega'^2,
\eea
where $F^i{}_{jkm}$ is the curvature of the manifold $N$ and prime denotes the derivative with respect
to $y$, that is,
$\omega'{}=\partial_y\omega$.
The non-zero components of the Ricci tensor and the scalar curvature
are
\bea
R_{00}&=& -2\alpha(\omega'^2-\omega''),
\\
R_{ij} &=& F_{ij}-h_{ij}e^{-2\omega}
\left[2\alpha\omega'^2
-\omega''\right],
\\
R&=&e^{2\omega}F
+4\alpha\omega''
-2\alpha(2\alpha+1)\omega'^2,
\label{226iga}
\eea
where $F_{ij}$ and $F$ is the Ricci tensor and the scalar
curvature of the manifold $N$.
In the case when the manifold $N$ is a sphere $S^{n-1}$,
the curvature is
\be
R{}^{ij}{}_{km} = \left(\delta^i{}_k \delta^j{}_{m}-\delta^i{}_m \delta^j_{k}
\right)
\left(\frac{1}{a^2}e^{2\omega}-\omega'^2\right).
\ee

In the case of the hyperbolic cylinder
with the warping function (for $y>0$)
\be
f(y)=\sinh(y/a),
\ee
that is,
\be
\omega(y)=-\log\left| \sinh(y/a)\right|,
\ee
we have
\bea
\omega' &=& -\frac{1}{a}\coth(y/a),
\\
\omega'' &=& \frac{1}{a^2\sinh^2(y/a)}.
\eea
Therefore, if the compact manifold $N$ is just the sphere
$S^{n-1}$ of radius $a$,
then the curvature of the manifold $M$ is constant
\bea
R{}^{0k}{}_{0i} &=&-\frac{1}{a^2}\delta^{k}{}_i,
\\
R{}^{ij}{}_{km} &=& -\frac{1}{a^2}
\left(\delta^i{}_k \delta^j{}_{m}-\delta^i{}_m \delta^j_{k}
\right).
\eea
Locally, this is nothing but the curvature of the hyperbolic
space $H^n=\Sigma\times_f S^{n-1}$.
Of course, since the warping function $f(y)=\sinh(y/a)$
is vanishing at $y=0$, the warped product manifold $M=\Sigma\times_f N$
is singular at $y=0$;
we actually have two copies of
the hyperbolic space $H^n$, one for $y>0$ and another for $y<0$.

Similarly, in the case of the cusp with the
warping function (for $y>0$)
\be
f(y)=\exp(-y/a)
\ee
that is,
\be
\omega(y)=\frac{y}{a},
\ee
the curvature has the form
\bea
R{}^{0k}{}_{0i} &=&
-\frac{1}{a^2}\delta^{k}{}_i,
\\
R{}^{ij}{}_{km} &=& e^{2y/a}F^{ij}{}_{km}
-\frac{1}{a^2}
\left(\delta^i{}_k \delta^j{}_{m}-\delta^i{}_m \delta^j_{k}
\right),
\eea


We will study in detail the
warping function
\be
f(y)=\frac{1}{\left[\cosh({} y/b)\right]^{\nu/\alpha}},
\ee
that is,
\be
\omega(y)=\frac{\nu}{\alpha}\log\cosh({} y/b);
\label{212iga}
\ee
with some parameter $\nu$.
This function behaves like cusps at infinity,
$|y|\to \infty$
\be
\omega(y)\sim \frac{\nu |y|}{\alpha b}
\ee
In this case
we have
\bea
\omega' &=& \frac{\nu}{\alpha b}{}\tanh({} y/b),
\\
\omega'' &=& \frac{\nu}{\alpha b^2}\frac{1{}}{\cosh^2({} y/b)},
\eea
and the curvature tensor is
\bea
R{}^{0k}{}_{0i} &=&
-\delta^{k}{}_i \frac{\nu^2}{\alpha^2 b^2}
\left[1
-\left(1+\frac{\alpha}{\nu}
\right)\frac{1}{\cosh^2(y/b)}\right],
\\
R{}^{ij}{}_{km} &=& \left[\cosh(y/b)\right]^{2\nu/\alpha}F^{ij}{}_{km}
-\left(\delta^i{}_k \delta^j{}_{m}-\delta^i{}_m \delta^j_{k}
\right)
\frac{\nu^2}{\alpha^2 b^2}\frac{1}{\cosh^2(y/b)}.
\eea
In a particular case when the compact manifold $N$ is flat,
(e.g. a torus $N=T^{n-1}$)
with zero curvature,
$F_{ijkl}=0$,
(or if the parameter $\nu$ is negative)
then the only non-zero components of the
curvature of the manifold $M$
at infinity approaches a constant
\be
R{}^{0k}{}_{0i} \sim
-\delta^{k}{}_i \frac{\nu^2}{\alpha^2 b^2}.
\ee
If the parameter $\nu$ is negative, then we have a funnel.
In the case of positive parameter $\nu$
the manifold has two cusps
$M_+=[a,\infty)\times_f N$ and
$M_-=(-\infty,-a]\times_f N$
with a finite volume determined by
(\ref{28iga}) with
\cite{prudnikov98}
\be
\beta=\int\limits_{-\infty}^\infty dy
\frac{1}{\cosh^{2\nu{}}({} y/b)}
= \sqrt{\pi}\frac{ \Gamma\left( \nu \right)}{\Gamma\left( \nu+\frac{1}{2} \right)}
b.
\label{241iga}
\ee



The equations of geodesics on the manifold $M$ are
\bea
\ddot y +\omega'e^{-2\omega}|\dot {\hat x}|^2
&=& 0,
\label{214iga}\\
\frac{D^2 \hat x^i}{ds^2}
&=&2\omega'\dot y \dot {\hat x}^i,
\label{215iga}\eea
where the dot denotes the derivative with respect to the
natural parameter $s$,
\be
|\dot {\hat x}|^2=h_{ij}\dot {\hat x}^i\dot {\hat x}^j,
\ee
and
\be
\frac{D^2 \hat x^i}{ds^2}
=\ddot {\hat x}^i
+\gamma^i{}_{jk}\dot {\hat x}^j\dot {\hat x}^k,
\ee
with $\gamma^i{}_{jk}$ being the Christoffel symbols of the metric $h$.
Obviously, the curves
\be
y(s)=s, \qquad \hat x^i(s)=\hat x'^i,
\ee
are geodesics.
More generally,
these equations have the integral
\be
\dot y^2+e^{-2\omega}|\dot {\hat x}|^2
=1.
\ee
Therefore, the first equation
(\ref{214iga}) takes the form
\be
\ddot y =\omega'(\dot y^2-1),
\ee
which can be integrated to get
\be
\dot y^2=1-c^2_1 e^{2\omega}.
\ee
with an integration constant $c_1$, that is,
\be
|\dot {\hat x}|^2=c^2_1e^{4\omega}.
\ee
This gives $y(s)$
implicitly,
\be
s=\int\limits_{y'}^{y(s)}\frac{dy''}{\sqrt{1-c^2_1 e^{2\omega(y'')}}}.
\ee
In particular, in the case of cusps,
when $\omega(y)=y/a$ (for $y>0$), we get
\bea
y(s)=-a\log\left\{
c_1\cosh\left[\frac{s}{a}
+\cosh^{-1}\left(\frac{1}{c_1}e^{-y'/a}\right)
\right]
\right\}
\eea

The second equation (\ref{215iga}) becomes
\be
\frac{D^2 \hat x^i}{ds^2}
=2\omega'\sqrt{1-c^2_1 e^{2\omega}}\; \dot {\hat x}^i.
\ee
As a consequence,  eq. (\ref{215iga}) gives
\be
h_{ij}\dot{\hat x}^j\frac{D^2 \hat x^i}{ds^2}
=2c^2_1e^{4\omega}\omega'\sqrt{1-c^2_1 e^{2\omega}}.
\ee

Let $d(\hat x,\hat x')$ be the geodesic distance
between the points $\hat x$ and $\hat x'$
on the manifold $N$
and $\sigma(\hat x,\hat x')$
be the Ruse-Synge function
equal to one half of the square of
the geodesic distance,
\be
\sigma(\hat x,\hat x')=\frac{1}{2}
\left[d(\hat x,\hat x')\right]^2.
\ee
It is determined by the equation
\cite{avramidi15}
\be
\sigma=\frac{1}{2}h^{ij}
\hat\partial_i\sigma\hat\partial_j\sigma
\ee
with the conditions
\be
\sigma(\hat x,\hat x)=0,
\qquad
\hat\partial_i\sigma(\hat x,\hat x')\Big|_{\hat x=\hat x'}=0;
\ee
Here $\hat\partial_i$ denote the partial derivatives with respect to
the coordinates $\hat x^i$.
Let
$d(y,\hat x,y',\hat x')$
be the the geodesic distance on the manifold $M$
between the points $(y,\hat x)$ and $(y',\hat x')$
and
\be
\rho(y,\hat x,y',\hat x')=\frac{1}{2}[d(y,\hat x,y',\hat x')]^2,
\ee
 be the corresponding Ruse-Synge function
in the manifold $M$.
It is determined by the equation
\be
\rho=\frac{1}{2}\left\{(\partial_y\rho)^2
+e^{2\omega}h^{ij}
\hat\partial_i\rho\hat\partial_j\rho
\right\}
\ee
with the conditions
\be
\rho(y,\hat x,y,\hat x)=0,
\qquad
\partial_y\rho(y,\hat x,y',\hat x)\Big|_{y=y'}=0,
\qquad
\hat\partial_i\rho(y,\hat x,y,\hat x')\Big|_{\hat x=\hat x'}=0;
\ee
also
\be
\partial^2_y\rho(y,\hat x,y',\hat x)\Big|_{y=y'}=1,
\qquad
\hat\partial_i\partial_y
\rho(y,\hat x,y,\hat x')\Big|_{\hat x=\hat x'}=0,
\ee
\be
\hat\partial_i\hat\partial_j
\rho(y,\hat x,y,\hat x')\Big|_{\hat x=\hat x'}=
e^{-2\omega}h_{ij}.
\ee
It is easy to see that
if $\hat x=\hat x'$, then
\be
\rho(y,\hat x,y',\hat x)=\frac{1}{2}(y-y')^2,
\ee
If the function $\omega$ is constant, then the solution
of this equation is
\be
\rho=\frac{1}{2}(y-y')^2+e^{-2\omega}\sigma(\hat x,\hat x').
\ee
In general, we parameterize the function $\rho$ by
\be
\rho=\frac{a^2}{2}\left(
\cosh^{-1}\chi\right)^2;
\ee
then the function $\chi$ satisfies the equation
\be
(\partial_y\chi)^2+e^{2\omega}h^{ij}
\hat\partial_i\chi\hat\partial_j\chi
=\frac{1}{a^2}\left(\chi^2-1\right).
\ee
Further, we let
\be
\chi=
\cosh\left(\frac{y-y'}{a}\right)
+\frac{1}{a^2}\eta(y,\hat x,y',\hat x')
\sigma(\hat x,\hat x'),
\ee
with some function
$\eta(y,\hat x,y',\hat x')$.
In the case of cusps, when
\be
\omega(y)=\frac{y}{a},
\qquad (\mbox{for } y>0),
\ee
this function can be found exactly
\be
\eta=\exp\left(-\frac{y+y'}{a}\right),
\ee
that is, the geodesic distance
on the manifold $M$ is
\be
d(y,\hat x,y',\hat x')=a
\cosh^{-1}\left[
\cosh\left(\frac{y-y'}{a}\right)
+\frac{1}{a^2}
\exp\left(-\frac{y+y'}{a}\right)
\sigma(\hat x,\hat x')
\right].
\label{267iga}
\ee

\section{Heat Kernel}
\setcounter{equation}{0}


Let $\nabla$ be the Levi-Civita connection and
$\nabla^*$ be the formal adjoint to $\nabla$ defined using the Riemannian
metric. The scalar Laplacian
$\Delta_M: C^\infty(M)\to C^\infty(M)$ is a partial differential operator of
the form
\be
\Delta_M=-\nabla^*\nabla=g^{\mu\nu}\nabla_\mu\nabla_\nu,
\label{1ms}
\ee
which in local coordinates takes the form
\be
\Delta_M=
|g|^{-1/2}\partial_\mu |g|^{1/2}g^{\mu\nu}
\partial_\nu.
\ee
The Laplacian on the warped product manifold $M=\Sigma\times_f N$ is
\bea
\Delta_M &=&
\partial_y^2-2\alpha\omega'{}\partial_y+e^{2\omega}\Delta_N,
\eea
where $\Delta_N=h^{ij}\nabla_i\nabla_j$ is the Laplacian
on the manifold $N$.
It is easy to see that the Laplacian can be written in the form
\be
\Delta_M = -e^{\alpha\omega}Le^{-\alpha\omega },
\label{38iga}
\ee
where
\be
L=
D_0-e^{2\omega}\Delta_N,
\label{35iga}
\ee
and $D_0$ is the ordinary differential operator
on $L^2(\Sigma, dy)$
defined by
\bea
D_0 &=&
 -(\partial_y-\alpha\omega'{})
(\partial_y+\alpha\omega'{})
\nn\\
&=&-\partial^2_y{}+Q_0,
\label{39iga}
\eea
where
\be
Q_0=\alpha^2\omega'^2{}
-\alpha\omega''{}.
\label{37iga}
\ee


The heat kernel
$U_M(t;x,x')=\exp(t\Delta_M)\delta_M(x,x')$
of the Laplacian is defined by
requiring it to
satisfy the equation
\be
\left(\partial_t-\Delta_M\right)U_M(t;x,x')=0,
\ee
with the initial condition
\be
U_M(0;x,x')=\delta_M(x,x'),
\ee
where
\bea
\delta_M(x,x') &=& |g|^{-1/4}(x)|g|^{-1/4}(x')\delta(x-x')
\nn\\
 &=&
\exp\left\{\alpha\left[\omega(y)+\omega(y')\right]\right\}
\delta(y{}-y{}')
\delta_N(\hat x,\hat x'),
\label{43iga}
\eea
is the covariant delta function on the manifold $M$.



By using the intertwining property (\ref{38iga})
the heat semigroup of the Laplacian,
$\exp(t\Delta_M): C^\infty(M)\to C^\infty(M)$,
takes the form
\be
\exp(t\Delta_M) = e^{\alpha\omega}
\exp\left(-tL\right)
e^{-\alpha\omega}.
\label{47iga}
\ee
Therefore, by acting on the delta function (\ref{43iga})
we obtain the heat kernel on the manifold $M$
in terms of the heat kernel of the operator $L$
\be
U_M(t;y,\hat x,y',\hat x')
=
\exp\left\{\alpha\left[\omega(y{})+\omega(y{}')\right]\right\}
U(t;y,\hat x,y',\hat x'),
\ee
where the heat kernel $U(t;y,\hat x,y',\hat x')$
of the operator $L$ is defined by the equation
\be
\left(\partial_t+L\right)U(t;y,\hat x,y',\hat x')=0,
\ee
with the initial condition
\be
U(0;y,\hat x,y',\hat x')=\delta(y-y')\delta_N(\hat x,\hat x').
\ee


Let $\{\mu_k\}_{k=0}^\infty$ be the spectrum
of the negative Laplacian $-\Delta_N$
on the manifold $N$ and $T_k$ be the corresponding
orthogonal spectral projections.
The multiplicities of the eigenvalues $\mu_k$ are
\be
d_k=\dim\Ker(\Delta_N+\mu_k).
\ee
Since
the manifold $N$ is closed,
the first eigenvalue (which is simple with
multiplicity $d_0=1$) is zero,
\be
\mu_0=0,
\ee
with the constant eigenfunction
\be
\varphi_0(\hat x)=[\vol(N)]^{-1/2}.
\ee

For example, in the case of the sphere $S^{n-1}$ the spectrum
of the negative Laplacian is \cite{avramidi23}
\be
\mu_k(S^{n-1})=\frac{1}{a^2}k(k+n-2),
\qquad
k=0,1,2,\dots,
\ee
with the multiplicities $d_0=1$ and
\be
d_k=(2k+n-2)\frac{(k+n-3)!}{(n-2)!k!},
\qquad
k=1,2,\dots.
\ee
In the case of the torus $T^{n-1}$ the eigenvalues are
\be
\mu_{k_1\dots k_{n-1}}(T^{n-1})=\sum_{j=1}^{n-1}\frac{k_j^2}{a_j^2},
\ee
where $a_j$ are the radii of the circles and $k_j\in\ZZ$.



Let $U_N(t;\hat x,\hat x')=\exp(t\Delta_N)\delta_N(\hat x,\hat x')$
be the heat kernel of the
Laplacian on the manifold $N$
satisfying the equation
\be
\left(\partial_t-\Delta_N\right)U_N(t;\hat x,\hat x')=0,
\ee
with the initial condition
\be
U_N(0;\hat x,\hat x')=\delta_N(\hat x,\hat x'),
\ee
where
\be
\delta_N(\hat x,\hat x')=|h|^{-1/4}(\hat x)|h|^{-1/4}(\hat x')\delta(\hat x-\hat x')
\ee
is the covariant delta function on the
manifold $N$.

Let $P_N$ be the projection onto the orthogonal complement
of the kernel of the Laplacian and $\tilde\Delta_N$ be the reduced
Laplacian defined by
\be
\tilde\Delta_N=P_N\Delta_N P_N.
\ee
The spectral resolution of the heat kernel on the manifold $N$
is
\be
U_N(t;\hat x,\hat x')
=\frac{1}{\vol(N)}
+\tilde U_N(t;\hat x,\hat x'),
\ee
where
\be
\tilde U_N(t;\hat x,\hat x')
=\sum_{k=1}^\infty
 e^{-t\mu_k}
T_k(\hat x,\hat x')
\ee
is the heat kernel of the positive operator
$
-\tilde\Delta_N.
$



We use the spectral resolution of the Laplacian on the manifold $N$
to separate variables and to get the heat kernel of the operator $L$
\be
U(t;y,\hat x,y',\hat x')=\sum_{k=0}^\infty
U_k(t;y,y')T_k(\hat x,\hat x'),
\ee
where the functions $U_k(t;y,y')=
\exp(-tD_k)\delta(y-y')$ are the heat kernels
of the
ordinary differential operators
\be
D_k=D_0+\mu_ke^{2\omega},
\label{426iga}
\ee
satisfying the equations
\be
\left(\partial_t+D_k\right)U_k(t;y,y')=0
\ee
with the initial conditions
\be
U_k(0;y,y')=\delta(y-y').
\ee

We will find it useful to
separate the zero mode of the operator $\Delta_N$.
By using the projection $T_0(\hat x,\hat x')=[\vol(N)]^{-1}$
to the kernel of the operator $\Delta_N$ we get
\be
U(t;y,\hat x,y',\hat x')=\frac{1}{\vol(N)}U_0(t;y,y')
+\tilde U(t;y,\hat x,y',\hat x'),
\label{329iga}
\ee
where the function $U_0(t;y,y')=\exp(-tD_0)$
is the heat kernel of the operator $D_0$ and
\bea
\tilde U(t;y,\hat x,y',\hat x')
&=&\sum_{k=1}^\infty
U_k(t;y,y')T_k(\hat x,\hat x').
\eea
Notice that
\be
\tilde U(t;y,\hat x,y',\hat x')
=\exp(-t\tilde L)\delta(y-y')\delta_N(\hat x,\hat x')
\label{331iga}
\ee
is the heat kernel of the operator
\be
\tilde L = D_0 -e^{2\omega}\tilde\Delta_N.
\label{332iga}
\ee


Let
\be
\Phi(t,\lambda{};y,y'{})
=\exp\left[-t\left(D_0+\lambda e^{2\omega}\right)\right]
\delta(y{}-y')
\label{418iga}
\ee
be the analytic continuation of the heat kernel of the
operator $\left(D_0+\lambda e^{2\omega}\right)$
defined for $\Real \lambda\ge 0$ by
\be
\left(\partial_t
+D_0+\lambda e^{2\omega}\right)\Phi(t,\lambda;y,y')=0,
\ee
with the initial condition
\be
\Phi(0,\lambda{};y,y')
=\delta(y-y').
\ee
This function is analytic function of $\lambda$ with a cut along
the negative real axis.
Then, of course,
\be
U_k(t;y,y')=\Phi(t,\mu_k;y,y').
\label{336iga}
\ee
Therefore,
we can use the Cauchy formula to represent the heat kernel
in terms of the resolvent $\tilde G_N(\lambda)=(-\tilde\Delta_N-\lambda)^{-1}$
of the positive operator $\tilde\Delta_N$,
\be
\tilde U(t;y,\hat x,y',\hat x')
=\int\limits_{c-i\infty}^{c+i\infty}
\frac{d\lambda}{2\pi i}\; \tilde G_N(\lambda;\hat x,\hat x')
\Phi(t,\lambda;y,y'),
\label{337iga}
\ee
where $c$ is a non-negative constant less than $\mu_1$, that is,
$
0\le c<\mu_1.
$
Further, we can represent this resolvent in terms of the heat kernel
of the operator $\tilde\Delta_N$, for $0\le \Real \lambda<\mu_1$,
\be
\tilde G_N(\lambda;\hat x,\hat x')
=\int\limits_0^\infty d\tau\; e^{\tau\lambda{}}
\tilde U_N(\tau;\hat x,\hat x').
\ee
Thus, we obtain the heat kernel
\be
\tilde U(t;y,\hat x,y',\hat x')
=
\int\limits_0^\infty d\tau\;
\tilde U_N(\tau;\hat x,\hat x')
\int\limits_{c-i\infty}^{c+i\infty}
\frac{d\lambda}{2\pi i}\;
e^{\tau\lambda{}}
\Phi(t,\lambda;y,y').
\label{339iga}
\ee
This enables one to compute the
heat trace
\be
\Tr\exp(-t\tilde L)
=
\int\limits_0^\infty d\tau\;
V(t,\tau)\Tr\exp(\tau\tilde\Delta_N),
\ee
where
\be
V(t,\tau)=
\int\limits_{c-i\infty}^{c+i\infty}
\frac{d\lambda}{2\pi i}\;
e^{\tau\lambda}
\Tr\exp\left[-t\left(
D_0+\lambda e^{2\omega}\right)\right].
\label{339igax}
\ee


An alternative method consists of employing the
wave kernel of the operator $\tilde\Delta_N$,
\be
W(p;\hat x,\hat x')
=\exp\left(ip\sqrt{-\tilde \Delta_N}\right)
\delta_N(\hat x,\hat x').
\ee
We use the Fourier transform
\bea
f(A) &=&
\int\limits_{-\infty}^\infty
dp\;
\exp\left(ip\sqrt{A}\right)
f(-\partial_p^2)
\delta(p),
\eea
to obtain the heat semigroup of the operator
$\tilde L=D_0-e^{2\omega}\tilde\Delta_N$,
\bea
\exp(-t\tilde L)
&=&
\int\limits_{-\infty}^\infty dp
\exp\left(ip\sqrt{-\tilde\Delta_N}\right)
\exp\left[-t\left(D_0-e^{2\omega}\partial_p^2\right)
\right]
\delta(p).
\eea
This gives
the heat kernel of the operator $\tilde L$
in the form
\bea
\tilde U(t;y,\hat x,y',\hat x') &=&
\int\limits_{-\infty}^\infty dp\;
W(p;\hat x,\hat x')
K(t;y,p,y',0),
\eea
where
\be
K(t; y,p, y', p')=
\exp\left[-t\left(D_0-e^{2\omega}\partial_p^2\right)
\right]
\delta(y-y')
\delta(p-p').
\ee

In the case of cusps,
for large $y>0$, when the function $\omega$ is given by
$\omega(y)\sim y/a$,
the operator $-D_0+e^{2\omega}\partial_p^2$
approaches the
Laplacian on the hyperbolic plane
$H^2$ (with a constant potential)
with a well known
heat kernel \cite{avramidi15},
\bea
K(t; y,p, y', p')
&=&\frac{a}{2(2\pi t)^{3/2}}
\exp\left[-\left(\frac{1}{4}+\alpha^2\right)
\frac{t}{a^2}
\right]
\nn\\
&&\hspace{-2cm}\times
\int\limits_{d/a}^\infty
ds\frac{s}{\sqrt{\cosh s-\cosh(d/a)}}
\exp\left(-\frac{s^2a^2}{4t}\right),
\eea
where $d$ is the geodesic distance given by
a formula similar to (\ref{267iga}),
\be
d(y,p,y',p')=a
\cosh^{-1}\left[
\cosh\left(\frac{y-y'}{a}\right)
+\frac{1}{a^2}
\exp\left(-\frac{y+y'}{a}\right)
(p-p')^2
\right].
\label{267igax}
\ee
The asymptotics of this heat kernel as
$t\to 0$ is
\be
K(t; y, p, y', p')
\sim \frac{1}{4\pi t}
\left(\frac{d}{a\sinh (d/a)}\right)^{1/2}
\exp\left(-\frac{d^2}{4t}\right).
\ee
Note that
\be
d(y,p,y,0)=
\cosh^{-1}\left[
1+\exp\left(-2\frac{y}{a}\right)
\frac{p^2}{a^2}
\right].
\label{267igay}
\ee

To compute the trace we need to regularize the wave
kernel as follows,
\bea
\Tr\exp(-t\tilde L) &=&
\int\limits_{0}^\infty dp\;
\int\limits_\RR dy
K(t;y,p,y,0)
\\
&&\times
\left\{
\Tr\exp\left(ie^{i\varepsilon}p\sqrt{-\tilde\Delta_N}
\right)
+\Tr\exp\left(-ie^{-i\varepsilon}p\sqrt{-\tilde\Delta_N}
\right)
\right\}.
\nn\eea



\section{Heat Trace on Compact Manifolds}
\setcounter{equation}{0}

The Laplacian $\Delta_N$ on the manifold $N$
has a discrete non-negative real spectrum
with finite multiplicities bounded from below
\cite{gilkey95}.
The spectrum is analyzed by
studying the heat trace
\be
\Tr\exp(t\Delta_N)=\dim\Ker\Delta_N
+\Tr\exp(t\tilde\Delta_N).
\ee
It is well known that it has the
asymptotic expansion as $t\to 0$,
\cite{gilkey95},
\be
\Tr\exp(t\Delta_N)
\sim (4\pi )^{-\alpha}\sum_{k=0}^\infty \frac{t^{k-\alpha}}{k!}A_k(N),
\label{11iga}
\ee
where
$A_k(N)$ are the spectral invariants called the global
heat kernel coefficients.
These coefficients play an important role in spectral geometry and
quantum field theory. They have been computed explicitly
up to $A_4$ \cite{gilkey95,avramidi91,avramidi00}.
Let $\varkappa$ be the largest positive integer
smaller than $\alpha=\frac{n-1}{2}$,
\be
\varkappa=\left[\frac{n}{2}\right]-1;
\ee
that is, if $n$ is odd then $\varkappa=\frac{n-3}{2}=\alpha-1$, and if $n$ is even
then $\varkappa=\frac{n-2}{2}=\alpha-\frac{1}{2}$.
Then
\be
\Tr\exp(t\Delta_N) = \Theta_{N,-}(t)
+\Theta_{N,+}(t),
\label{16igaxx}
\ee
where the function
\be
\Theta_{N,-}(t) = (4\pi)^{-\alpha}\sum_{k=0}^{\varkappa}
 \frac{1}{k!}A_k(N)t^{k-\alpha},
\label{122iga}
\ee
contains
only negative powers of $t$ and the function
\be
\Theta_{N,+}(t) \sim
(4\pi)^{-\alpha}\sum_{k=\varkappa+1}^{\infty}
\frac{1}{k!}A_k(N)t^{k-\alpha{}},
\ee
contains only non-negative powers of $t$.

The zeta function is defined by a modified
Mellin transform of the heat trace,
\bea
\zeta_N(s)
=\Tr(-\tilde\Delta_N)^{-s}
&=&\frac{1}{\Gamma(s)}\int\limits_0^\infty dt\;
t^{s-1}\Tr\exp(t\tilde\Delta_N).
\label{12iga}
\eea
This function is analytic for
$\Real s>\alpha{}$ and has a meromorphic analytic continuation.
The analytic structure of the zeta function is exhibited by
the representation \cite{avramidi91,avramidi00,avramidi10},
\be
\zeta_N(s)=\frac{\Gamma(s-\alpha{})}{\Gamma(s)}Z_N(s),
\label{18iga}
\ee
where $Z_N(s)$ is an entire function.
The entire function  $Z_N(s)$ is expressed in terms of the heat trace by
\be
Z_N(s)=\frac{1}{\Gamma\left(s-\alpha{}+k\right)}
\int\limits_0^\infty dt\; t^{s-1-\alpha{}+k}
(-\partial_t)^k\left[t^{\alpha{}}\Tr\exp(t\tilde\Delta_N)\right],
\ee
where $k$ is a sufficiently large positive integer, $k>\alpha{}$.

It is easy to see that  the values at the
points $s=\alpha{}-k$, $k=0,1,2,\dots$, are
\be
Z_N\left(\alpha{}-k\right)=(-1)^{k}(4\pi)^{-\alpha{}}A_{k}(N),
\label{110iga}
\ee
where $A_k(N)$ are the heat trace coefficients of the manifold $N$
defined by (\ref{11iga}).
Therefore, if $n$ is even (that is, $\alpha$ is half-integer) then the zeta function
has an infinite number of
simple poles at the half-integer points
$s_k=\alpha{}-k$, $k=0,1,2,\dots,$ with the residues
\be
\Res\zeta_N\left(\alpha{}-k\right)
=\frac{1}{k!\Gamma\left(\alpha{}-k\right)}
(4\pi)^{-\alpha{}}A_{k}(N).
\ee
If $n$ is odd (that is, $\alpha$ is an integer),
then the zeta function has only a finite number
of these poles at the integer points
$1,2, \dots, \alpha-1,
\alpha{}$.

In any case, the zeta function is analytic at $s=0$.
The value at $s=0$ gives the regularized number of modes;
it is easy to see that
\be
\zeta_N(0)=
\left\{
\begin{array}{ll}
0, & \mbox{ for even } n,
\\[10pt]
\displaystyle
\frac{(4\pi)^{-\alpha{}}}{\Gamma\left(\alpha{}+1\right)}A_{\alpha{}}(N),
& \mbox{ for odd } n.
\end{array}
\right.
\ee

The value of the derivative at $s=0$ defines the regularized
determinant of the operator
\be
\Det(-\Delta_N)=\exp\left\{-\zeta'_N(0)\right\}.
\ee
It is given by the integral:
for even $n=2j$,
\be
\zeta'_N(0)=\frac{\sqrt{\pi}}{\Gamma(j+\frac{1}{2})}
\int\limits_0^\infty dt\; t^{-\frac{1}{2}}
(-\partial_t)^{j}\left[t^{j-\frac{1}{2}}\Tr\exp(t\tilde\Delta_N)\right],
\ee
and
for odd $n=2j-1$,
\bea
\zeta'_N(0) &=& \frac{(-1)^{j-1}}{(j-1)!}\int\limits_0^\infty dt\;
\left\{
\log t
+\psi(j)
\right\}
(-\partial_t)^j\left[t^{j-1}\Tr\exp(t\tilde\Delta_N)\right],
\eea
where $\psi(z)=\Gamma'(z)/\Gamma(z)$ is the logarithmic derivative
of the gamma function.

The heat trace can be obtained from the zeta function by the inverse Mellin
transform
\be
\Tr\exp(t\tilde\Delta_N)=
\int\limits_{\sigma-i\infty}^{\sigma+i\infty}
\frac{ds}{2\pi i}\;t^{-s}\;
\Gamma\left(s-\alpha\right)Z_N\left(s\right),
\label{16iga}
\ee
where $\sigma$ is a positive constant,
$\sigma> \alpha$.
More generally,
\bea
(-\partial_t)^k\left[
t^\alpha\Tr\exp(t\tilde\Delta_N)
\right]
=
\int\limits_{\tilde\sigma-i\infty}^{\tilde\sigma+i\infty}
\frac{ds}{2\pi i}\;t^{-s+\alpha-k}\;
\Gamma\left(s-\alpha+k\right)
Z_N\left(s\right),
\eea
where $\tilde\sigma>\alpha-k$.

Notice that this formula can also be used for a complex
variable $t$ with a cut along the negative real axis.
By moving the contour of integration to the left and
using eq. (\ref{110iga}) we obtain a
more general formula
\be
\Tr\exp(t\tilde\Delta_N) = \Theta_{N,-}(t)
+\Theta_{N,+}(t),
\label{16igaxy}
\ee
where the function $\Theta_{N,-}(t)$ is defined in (\ref{122iga})
and
\bea
\Theta_{N,+}(t) &=&
\int\limits_{\sigma_\varkappa-i\infty}^{\sigma_\varkappa+i\infty}
\frac{ds}{2\pi i}\;t^{-s}\;
\Gamma\left(s-\alpha\right)Z_N\left(s\right),
\eea
with  $\alpha-\varkappa-1<\sigma_\varkappa<\alpha-\varkappa$.
Similarly, we have
\bea
(-\partial_t)^k\left[
t^\alpha\Theta_{N,+}(t)
\right]
=
\int\limits_{\tilde\sigma-i\infty}^{\tilde\sigma+i\infty}
\frac{ds}{2\pi i}\;t^{-s+\alpha-k}\;
\Gamma\left(s-\alpha+k\right)
Z_N\left(s\right),
\eea
where $\tilde\sigma>\alpha-k$.

By taking the derivative, this equation also gives the traces
for any positive integer $j>0$,
\bea
\Tr\left\{(-\Delta_N)^j\exp(t\Delta_N)\right\}
& =&
(4\pi)^{-\alpha{}}\sum_{k=0}^{\varkappa} \frac{1}{k!}
\frac{\Gamma\left(\alpha{}+j-k\right)}
{\Gamma\left(\alpha{}-k\right)}
A_k(N)t^{k-j-\alpha{}}
\nn
\\
&&
\hspace{-3cm}
+\int\limits_{\sigma_\varkappa-i\infty}^{\sigma_\varkappa+i\infty}
\frac{ds}{2\pi i}\;t^{-s-j}\;
\frac{\Gamma\left(s+j\right)}
{\Gamma\left(s\right)}
\Gamma\left(s-\alpha\right)Z_N\left(s\right),
\label{16igax}
\eea


Note that we can also define the trace
\be
\Tr(-z\tilde\Delta_N)^{-s}
=z^{-s}\zeta_N(s)
\ee
for a complex variable $z$ with a cut along the negative real axis.
In particular,
\be
\Tr(i\tilde\Delta_N)^{-s}=e^{-is\pi/2}\zeta_N(s).
\ee
More generally, for $\Real \lambda<0$ and $\Real s> \alpha{} $
we can define the trace
\bea
\Tr(-z\Delta_N-\lambda)^{-s}
&=&
(-\lambda)^{-s}\dim\Ker\Delta_N
\\
&&
+\frac{(-\lambda)^{-s}}{\Gamma(s)}\int\limits_{\sigma-i\infty}^{\sigma+i\infty}
\frac{dw}{2\pi i}
\left(\frac{-\lambda}{z}\right)^{w}
\Gamma(w-\alpha)\Gamma(s-w)Z_N\left(w\right),
\nn\eea
where $\alpha<\sigma<\Real s$.

\section{One Dimensional Heat Kernel}
\setcounter{equation}{0}

Let $D$ be a Schr\"odinger operator acting on smooth functions
on $\RR$ of the form
\be
D=-\partial_y^2+Q(y),
\ee
where the potential $Q(y)$ is an even function.
Let $U(t;y,y')$ be the heat kernel of the operator $D$
 satisfying
the equation
\be
\left(\partial_t-\partial_y^2
+Q(y)\right)
 U(t;y,y')=0
\ee
with the initial condition
\be
U(0;y,y')=\delta(y-y').
\ee

The asymptotics of the heat kernel diagonal
$U(t;y,y)$ as $t\to 0$
has the following form
\cite{avramidi00b}
\be
U(t;y,y)\sim
(4\pi)^{-1/2}\sum_{k=0}^\infty
\frac{t^{k-1/2}}{k!}c_k(y),
\label{56iga}\ee
where the coefficients $c_k$
are differential polynomials of the potential $Q$
of degree $k$,
\be
c_k=\sum_{j=1}^k\sum_{|{\bf m}|=2k-2j}
(-1)^j
c({\bf m})Q^{(m_j)}\cdots Q^{(m_1)},
\ee
where ${\bf m}=(m_1,\dots m_j)$ is a multiindex,
$|{\bf m}|=m_1+\cdots+m_j$, and the coefficients
$c({\bf m})$ are computed explicitly
in \cite{avramidi00b}.
In particular,
\bea
c_0 &=& 1,
\\
c_1 &=& -Q,
\\
c_2 &=& Q^2-\frac{1}{3}Q''.
\eea

One can show that the heat kernel diagonal $U(t;y,y)$
satisfies the equation
\cite{avramidi14,avramidi95,avramidi15}
\be
\left(\partial_t\partial_y-\frac{E}{4}\right)U(t;y,y)
=0,
\ee
where $E$ is a third-order differential operator
\bea
E &=& \partial_y^3-2Q\partial_y-2\partial_y Q.
\eea
Therefore, the coefficients $c_k$
satisfy
the recurrence relations
\be
\partial_y c_k=\frac{k}{2(2k-1)}Ec_{k-1},
\qquad
k\ge 1,
\ee
with the initial condition
\be
c_0=1.
\ee
The formal
solution of this recurrence is, for $k\ge 1$,
\be
c_k=\frac{(k!)^2}{(2k)!}B^{k}\cdot 1,
\label{59iga}\ee
where
\be
B=\partial_y^{-1}E.
\ee
One can sum the asymptotic series formally to get
\bea
U(t;y,y)
&\sim&
(4\pi t)^{-1/2}\varphi\left(tB\right)\cdot 1
=(4\pi t)^{-1/2}\left[1-tf(tB)Q\right],
\eea
where
\bea
f(t) &=&
\sum_{k=0}^\infty
\frac{k!}{(2k+1)!}t^{k}
\nn\\
&=&\int\limits_0^1 ds\exp\left[
\frac{1}{4}(1-s^2)t
\right]
\eea
and
\bea
\varphi(t)&=&\sum_{k=0}^\infty
\frac{k!}{(2k)!}t^{k}
\nn\\
&=&
\left(1+2t\partial_t\right)
f(t).
\eea

By using this solution one can get the linear and quadratic terms
in all coefficients $c_k$; they have the form
\cite{avramidi14,avramidi15}
\be
c_k=\frac{k!(k-1)!}{(2k-1)!}
\left\{-\partial_y^{2(k-1)}Q
+(2k-1)Q\partial_y^{k-2}Q
\right\}
+O(\partial_y(QQ))+O(Q^3),
\label{518iga}\ee
where the neglected terms are of order higher than
or equal to three and the total derivatives.
Thus, it has the asymptotic expansion
\be
U(t;y,y)=(4\pi t)^{-1/2}
\left\{1-tf(t\partial_y^2)Q
+\frac{t^2}{2}Qf(t\partial_y^2)Q
+O(\partial_y(QQ))
+O(Q^3)\right\}.
\label{519iga}
\ee

On another hand one can separate the terms without
the derivatives of the potential
\be
c_k=(-1)^kQ^k+O(\partial Q).
\ee
These terms can also be summed up to get a restructured
asymptotic expansion
\be
U(t;y,y)\sim
(4\pi )^{-1/2}\exp(-tQ)\sum_{k=0}^\infty \frac{t^{k-1/2}}{k!}\tilde c_k,
\label{521iga}
\ee
where the coefficients $\tilde c_k$ are differential polynomials
in the potential $u$,
\be
\tilde c_k=\sum_{j=0}^k {k\choose j}Q^j c_{k-j}.
\ee


\section{Heat Kernel of the Operators $D_k$ with
$k\ge 1$}
\setcounter{equation}{0}

The operators $D_k$, $k\ge 1$, (\ref{426iga}),
are Schr\"odinger operators
\bea
D_k &=&
-\partial_y^2+Q_k(y),
\label{41iga}\eea
with the potentials
\be
Q_k(y)
=\alpha^2\omega'^2{}-\alpha\omega''{}+\mu_k e^{2\omega}.
\label{42iga}
\ee

In the particular
case of cusps (\ref{212iga})
the potentials $Q_k$, (\ref{42iga}),
of the operator $D_k$ defined by (\ref{41iga})
are
\be
Q_k= \frac{1}{b^2}\left\{\nu^2
-\frac{\nu{}(\nu{}+1){}}{\cosh^2({} y/b)}
\right\}
+\mu_k [\cosh(y/b)]^{2\nu/\alpha}.
\label{52iga}
\ee
By introducing a new variable
\be
z=\tanh({} y/b)
\ee
the operators take the form
\be
D_k=\frac{1}{b^2}
\left\{-(1-z^2)^2\partial_z^2+2z(1-z^2)\partial_z
-\nu{}(\nu{}+1)(1-z^2)
+\frac{\mu_k b^2}{(1-z^2)^{\nu/\alpha}}
+\nu^2
\right\}.
\ee
This potential is symmetric, has a minimum at $y=0$, and behaves like a harmonic oscillator as $y\to 0$
\be
Q_k=\frac{1}{b^2}(\mu_k b^2-\nu)
+\frac{1}{b^4}\left[
\frac{\nu}{\alpha}\mu_k b^2+ \nu(\nu+1)
\right]y^2+O(y^4)
\label{52igaxy}
\ee
and, for $k\ge 1$, when $\mu_k>0$, grows at infinity
$|y|\to\infty$,
\be
Q_k=\frac{\mu^k}{2^{2\nu/\alpha}} \exp\left(
2\frac{\nu }{\alpha b}|y|\right)
+O\left(\exp\left(
2\frac{\nu-\alpha }{\alpha b}|y|\right)\right).
\label{77igax}\ee

Thus the operators $D_k$
with $k\ge 1$, when $\mu_k>0$, have confining
potential $Q_k$ going to infinity at infinity
and, therefore,
are strictly positive.
These operators are self-adjoint by construction and
are non-negative
since all eigenvalues $\mu_k$ are positive,
$\mu_k> 0$,
\be
(\varphi,D_k\varphi)=||(\partial_y+\alpha\omega')\varphi||^2
+\mu_k(\varphi, e^{2\omega}\varphi)\ge 0,
\ee
so, they do not have any zero modes, that is,
\be
\dim\Ker D_k=0 \qquad\mbox{for }\ k\ge 1.
\ee

On the circle, $\Sigma=S^1$, the spectrum of all operators $D_k$ is
discrete. For the noncompact case, $\Sigma=\RR$, the nature of the spectrum
depends on the behavior of the potentials $Q_k$ at infinity, as $|y|\to \infty$.
By assumption the function $\exp(2\omega(y))$ goes to infinity at infinity.
They have discrete positive real spectrum consisting
of simple eigenvalues $\lambda_{k,j}>0$, $j=1,2,\dots$.
Therefore, for $\mu_k>0$ the spectrum of the operators $D_k$,
with $k\ge 1$, is discrete.
Let $\lambda_{k,j}=\lambda_j(D_k)$, $j\in\ZZ_+$, be the eigenvalues of the operator
$D_k$, with $k\ge 1$, and $P_{k,j}=P_j(D_k)$ be the
corresponding orthogonal spectral projections to the eigenspaces.
Then the heat kernel of the operator $D_k$ with $k\ge 1$ is
\be
U_k(t;y,y')=\sum_{j=1}^\infty
\exp\left(-t\lambda_{k,j}\right)P_{k,j}(y,y').
\label{16zz}
\ee
Notice that since $\lambda_{k,j}>0$ for all $k,j\ge 1$, these
heat kernels decrease exponentially as $t\to \infty$,
\be
U_k(t;y,y')\sim e^{-t\lambda_{k,1}}P_{k,1}(y,y')
+\cdots .
\ee


Since the operators $D_k$ with $k\ge 1$ have a discrete
unbounded spectrum, $\{\lambda_{k,j}\}_{j=1}^\infty$,
the heat trace of these operators is well defined,
\be
\Tr \exp(-tD_k)
=\sum_{j=1}^\infty d_{k,j}e^{-t\lambda_{k,j}},
\label{411iga}
\ee
where
\be
d_{k,j}=\dim\Ker(D_k-\lambda_{k,j})
\ee
is the multiplicity of the eigenvalue $\lambda_{k,j}$.
Since the spectrum of the operator $D_k$, $k\ge 1$, is discrete,
there is a well defined asymptotic expansion of the heat trace
as $t\to 0$. The asymptotic expansion of the heat kernel
is given by (\ref{521iga}),
\be
U_k(t;y,y)\sim
(4\pi)^{-1/2}
\exp\left[-tQ_k(y)\right]
\sum_{j=0}^\infty \frac{t^{j-1/2}}{j!}b_{k,j}(y),
\label{91igax}
\ee
where $b_{k,j}$ are differential polynomials in the potential $Q_k$
of degree $j$ with initial condition
$b_{k,0}=1$.
Therefore, the heat trace asymptotic expansion is determined
by
\be
\Tr\exp(-tD_k)\sim
(4\pi )^{-1/2}
\sum_{j=0}^\infty \frac{t^{j-1/2}}{j!}B_{k,j}(t),
\label{91iga}
\ee
where
\be
B_{k,j}(t)=
\int\limits_\RR dy\;
\exp\left[-tQ_k(y)\right]b_{k,j}(y).
\label{91igaz}
\ee

The leading asymptotics
is determined by the behavior of the potential $Q_k(y)$ at infinity.
If, as $|y|\to\infty$
\be
Q_k(y)\sim \mu_k\left(\frac{|y|}{b}\right)^q
\ee
with some positive $q>0$, then
\be
\Tr\exp(-tD_k)\sim
\frac{b}{2\pi}\Gamma\left(\frac{1}{q}\right)
\mu_k^{-1/q}t^{-1-1/q}+\cdots.
\ee

In the case of the cusps, when the potential has the form
(\ref{52iga}) with the asymptotic (\ref{77igax}),
the main asymptotics is
\be
\Tr\exp(-tD_k)=
-b(4\pi t)^{-1/2}
\left\{
\frac{\alpha}{\nu}[\log(\mu_k t)-\gamma]
-2+O(t)\right\},
\label{418igax}
\ee
where $\gamma=0.577...$ is the Euler constant.

\section{Heat Kernel of the Operator $D_0$}
\setcounter{equation}{0}

The operator $D_0$, defined by (\ref{39iga}),
is special.
This operator  is self-adjoint by construction and
is non-negative,
\be
(\varphi,D_0\varphi)
=||(\partial_y+\alpha\omega')\varphi||^2.
\ee
The potential $Q_0$
of the operator $D_0$
does not have the exponential term and is given by
(\ref{37iga}).
The kernel of this operator is defined by
the normalized solutions of the equation
\be
D_0\psi=0.
\ee
This equation
has  two solutions
\be
\psi^{-}_{0,1}(y)=c^{-}_{0,1}e^{-\alpha\omega(y)},
\qquad
\psi^{+}_{0,1}(y)=c^{+}_{0,1}e^{-\alpha\omega(y)}
\int_0^y dy' e^{2\alpha\omega(y')}.
\ee
Of course, they are normalizable on the circle $\Sigma=S^1$.
The condition that the function $\psi^{-}_{0,1}$
 is normalizable in the noncompact case,
$\Sigma=\RR$,
is exactly the same
as the condition of the finite volume of the noncompact manifold $M$.
Since the function $\omega$ grows at infinity, then the other solution,
$\psi^{+}_{0,1}$, grows at infinity and is not normalizable.
Therefore, in the noncompact case the operator $D_0$ has
only one zero eigenvalue $\lambda_{0,1}=0$.
Therefore,
\be
\dim\Ker D_0 =
\left\{
\begin{array}{ll}
 2, &\mbox{on}\ S^1,
\\
1, & \mbox{on}\ \RR.
\end{array}
\right.
\ee


On the circle $S^1$ the spectrum of the operator $D_0$ is discrete.
In the noncompact case, $\Sigma=\RR$,
depending on the behavior
of the function $\omega(y)$ at infinity, the spectrum of the operator
$D_0$ could be both
discrete and continuous.
For example, if $\omega(y)=y/b$, then the operator
has the form $D_0=-\partial_y^2+\alpha^2/b^2$
with a continuous spectrum.
On another hand, if $\omega(y)=(y/b)^2$ then
the operator takes the form
of a harmonic oscillator,
\be
D_0=-\partial_y^2+\frac{4\alpha^2 }{b^4}y^2-\frac{2\alpha}{b^2},
\ee
with
the discrete spectrum
\be
\lambda_{0,j}=\frac{4\alpha}{b^2} j,
\qquad j=0,1,2,\dots.
\ee
More generally, if the function $\omega(y)$
grows at infinity like $|y|^p$ with $p\ge 1$ then the potential
grows at infinity like $|y|^{2(p-1)}$.
Therefore, if $p>2$, it grows faster than $y^2$ and
the spectrum of the operator $D_0$
is discrete, and if $p<3/2$ then it grows at infinity slower than
$|y|$, then the spectrum is continuous (it could still have a finite
number of discrete eigenvalues).

As $t\to \infty$ the heat kernel $U_0$ behaves like
\be
U_0(t;y,y')\sim P_0(y,y')+\cdots,
\ee
where $P_0(y,y')$ is the projection onto the kernel
(which is two-dimensional in the compact case of $S^1$ and
one-dimensional in the noncompoact case of $\RR$).


We can construct the heat kernel $U_0$ more directly in terms of the resolvent
of the operator $D_0$ by
\be
U_0(t;y,y')=\int\limits_{c-i\infty}^{c+i\infty}
\frac{d\lambda}{2\pi i}
e^{-t\lambda}G_0(\lambda;y,y'),
\label{617iga}
\ee
where $c$ is a negative constant and $G_0$ is the resolvent
defined by  the equation
\be
(D_0-\lambda)G_0(\lambda;y,y')=\delta(y-y').
\ee
The resolvent is defined first in the region where $\lambda$ is
a sufficiently large negative real parameter
and then analytically continued
to the whole complex plane. It will have some singularities on the positive
real line depending on the spectrum of the operator $D_0$.
It is constructed as follows.

Let $\tilde G_0(\lambda;y,y')$
be a function that satisfies the homogeneous equation
in both variables
\be
(D_{0,y}-\lambda)\tilde G_0(\lambda;y,y')=0,
\qquad
(D_{0,y'}-\lambda)\tilde G_0(\lambda;y,y')=0,
\ee
and satisfies the asymptotic conditions
\be
\lim_{y\to -\infty}\tilde G_0(\lambda;y,y')=0,
\qquad
\lim_{y'\to +\infty}\tilde G_0(\lambda;y,y')=0.
\ee
Then the resolvent kernel of the operator $L$ has the form
\bea
G_0(\lambda;y,y')
&=&{1\over C(\lambda)}\Bigl\{\theta (y'-y)\tilde G_0(\lambda;y,y')
+\theta (y-y')\tilde G_0(\lambda;y',y)
\Bigr\},
\label{519igax}
\eea
where $\theta(x)$ is the Heaviside step function and
$C(\lambda)$ is a constant defined by
\be
C(\lambda)=\left[
\partial_y \tilde G_0(\lambda;y,y')
-\partial_{y'} \tilde G_0(\lambda;y,y')
\right]\Big|_{y=y'}.
\ee

The spectrum is determined by the singularities of the resolvent.
Since the operator $D_0$ is self-adjoint and non-negative, then
all singularities of the resolvent are on the positive real axis.
For the continuous spectrum
the resolvent has the branch cut singularity along the positive
real axis. If the resolvent is a meromorphic function
then the spectrum is discrete. If it has branch singularities
then there a continuous spectrum.


In the case of cusps, the potential
has the form
(so called P\"oschl-Teller potential)
\bea
Q_0 &=&
\frac{1}{b^2}\left\{\nu^2
-\frac{\nu(\nu+1)}{\cosh^2(y/b)}\right\}.
\label{525iga}
\eea
Such potentials enable the exact solution
\cite{derezinski11}.
This potential behaves like a harmonic oscillator as $y\to 0$
\be
Q_0=-\frac{\nu}{b^2}{}{}
+\frac{\nu{}(\nu{}+1)}{b^4}
{} y^2
+O(y^4)
\ee
and approaches a constant at infinity as $|y|\to\infty$,
\be
Q_0=\frac{\nu^2}{b^2}
+O\left(e^{-2{} |y|/b}\right).
\label{77iga}\ee

The operator $D_0$ has a more complicated
spectrum.
Of course, it still has the simple zero eigenvalue
$\lambda_{0,0}=0$.
In the interval
$[0, \nu^2)$ it might
have some discrete eigenvalues
$
\lambda_{0,j}
$
with $j=1,\dots, N$, depending on the value of the parameter
$\nu$.
The rest of the spectrum, $[\nu^2,\infty)$, is continuous.

To study the structure of the spectrum we compute the resolvent.
We cut the complex plane $\lambda$ along the real axis
from $\nu^2$ to infinity and
define
\be
\mu=\sqrt{-\lambda b^2+\nu^2},
\ee
with $\Real\mu\ge 0$.
We also define another commonly used spectral parameter
$s$ such that
\be
\lambda=\frac{1}{b^2}(\nu^2-\mu^2)
=\frac{1}{b^2}s\left(2\nu-s\right)
\ee
by
\be
s=\nu+\mu,
\ee
so that $\Real s\ge \nu$.
This maps works as follows:
the whole comnplex plane of $\lambda$ with a cut along
the real axis from $\nu^2/b^2$ to infinity is mapped to the
right half-plane of $\mu$. The region $\Real \lambda>\nu^2/b^2$
is mapped to the region
\be
|\Imag\mu|>\Real\mu>0;
\ee
The left half-plane, $\Real \lambda<0$,
is mapped to the interior of the hyperbola
\be
\Real \mu >\sqrt{(\Imag\mu)^2+\nu^2-b^2\Real\lambda};
\ee
and the vertical strip $0<\Real\lambda<\nu^2/b^2$ is mapped
to the region
\be
|\Imag\mu|<\Real \mu <\sqrt{(\Imag\mu)^2+\nu^2-b^2\Real\lambda}.
\ee
Then the interval
$(-\infty,\nu^2/b^2]$ on the real axis in the complex plane
$\lambda$ is mapped to the positive real axis $[0,\infty)$
in the complex plane $\mu$.
The upper bank of the cut,
that is, $\lambda=(\nu^2+p^2)/b^2+i\varepsilon$, with positive $p>0$
is mapped to the negative imaginary half-axis of the
complex plane $\mu$,
$\mu=-ip$,
and the lower bank of the cut,
that is, $\lambda=(\nu^2+p^2)/b^2-i\varepsilon$,
is mapped to the positive imaginary half-axis of the,
$\mu=ip$.

The resolvent is determined by (\ref{519iga})
where the function $\tilde G_0(\lambda;y,y')$
satisfies the equation
\be
\left\{\partial_y^2
+\frac{\nu(\nu+1)}{b^2\cosh^2(y/b)}
-\frac{\mu^2}{b^2}\right\}\tilde G_0(\lambda;y,y')=0.
\ee
For  $y, y'\to \infty$
or $y,y'\to -\infty$ the resolvent has the well-known
form
\be
G_0(\lambda;y,y')\sim \frac{b}{2\mu{}}
\exp\left(-\mu{}|y-y'|/b\right).
\label{77igaxy}
\ee

In terms of the new variable
$
z=\tanh({} y/b)
$
we obtain the Legendre equation
\be
\left\{-(1-z^2)\partial_z^2+2z\partial_z
-\left[\nu{}(\nu{}+1)
-\frac{\mu^2}{(1-z^2)}
\right]\right\}\tilde G_0(\lambda;y,y')
=0.
\ee
The two linearly independent solutions of this equation are
given by the associated Legendre functions
$P^{\pm\mu}_\nu(\pm z)$
and $Q^{\pm \mu}_\nu(\pm z)$
\cite{erdelyi53}.
Any two linearly independent
solutions can be expressed in terms of the other two solutions.
We choose for the two solutions the functions
\bea
E_+(\mu,y)&=&
\frac{\Gamma(1+\mu+\nu)\Gamma(\mu-\nu)}{\Gamma(\mu)}
P^{-\mu}_\nu(-z),
\\
E_-(\mu,y)&=&
\frac{\Gamma(1+\mu+\nu)\Gamma(\mu-\nu)}{\Gamma(\mu)}
P^{-\mu}_\nu(z),
\eea
where
\bea
P^{-\mu}_{\nu{}}(z) &=& \frac{1}{\Gamma(1+\mu)}
\left(\frac{1-z}{1+z}\right)^{\mu/2}
F\left(-\nu{},\nu{}+1;1+\mu;\frac{1-z}{2}\right),
\\
P^{-\mu}_{\nu{}}(-z)
&=&\frac{1}{\Gamma(1+\mu)}
\left(\frac{1+z}{1-z}\right)^{\mu/2}
F\left(-\nu{},\nu{}+1;1+\mu;\frac{1+z}{2}\right),
\eea
are associated Legendre functions and
$F(a,b;c;z)$ denotes the hypergeometric function
\cite{erdelyi53}.
For integer values of $\nu$ the  functions
$F\left(-\nu{},\nu{}+1;1\pm\mu;\frac{1\pm z}{2}\right)$
are polynomials.
Notice that
\be
E_-(\mu,y)=E_+(\mu,-y).
\label{726igax}
\ee

These function can be analytically continued to meromorphic functions
of $\mu$.
They satisfy the equations
\be
\left(D_0+\frac{\mu^2}{b^2}\right)E_\pm(\mu,y)=0,
\ee
in fact, they satisfy the equations on the whole manifold $M$
\be
\left(\Delta_M-\frac{\mu^2}{b^2}\right)E_\pm(\mu,y)=0.
\ee

By using the relations
\cite{erdelyi53}
\bea
{}F{}(a,b;c;z)&=&(1-z)^{c-a-b}{}F{}(c-a,c-b;c;z)
\\[10pt]
F(a,b;c;1-z)&=&
\frac{\Gamma(c)\Gamma(c-a-b)}{\Gamma(c-a)\Gamma(c-b)}F(a,b;a+b-c+1;z)
\\
&&
+z^{c-a-b}\frac{\Gamma(c)\Gamma(a+b-c)}{\Gamma(a)\Gamma(b)}F(c-a,c-b;c-a-b+1;z)
\nonumber
\eea
we get
\bea
&&
-\frac{\pi}{\Gamma(1-\mu+\nu)\Gamma(-\mu-\nu)}
\frac{1}{\Gamma(1+\mu)}
F(-\nu,\nu+1;1+\mu;z)
\nn\\
&&
=\sin(\pi\mu)\frac{1}{\Gamma(1-\mu)}F(-\nu,\nu+1;1-\mu;1-z)
\nn\\
&&
+z^{-\mu}(1-z)^{\mu}
\sin(\pi\nu)\frac{1}{\Gamma(1-\mu)}
F(-\nu,\nu+1;1-\mu;z)
\label{726iga}
\eea
and, therefore,
\bea
&&
-\frac{\pi}{\Gamma(1-\mu+\nu)\Gamma(-\mu-\nu)}
P^{-\mu}_\nu(z)
=\sin(\pi\mu)
P^{\mu}_\nu(-z)
+\sin(\pi\nu)
P^{\mu}_\nu(z).
\label{725iga}
\eea


We study the behavior of these solutions at infinity
$y\to +\infty$, ($z\to 1$), and $y\to -\infty$, ($z\to -1$).
Notice that
\be
\frac{1+z}{1-z}=e^{2y/b}.
\ee
By using the hypergeometric series and the eq. (\ref{725iga})
in the form
\bea
&&
P^{-\mu}_\nu(z)
=
\frac{\Gamma(\mu)\Gamma(1-\mu)}{\Gamma(1+\mu+\nu)\Gamma(\mu-\nu)}
P^{\mu}_\nu(-z)
+
\frac{\sin(\pi\nu)}{\sin(\pi\mu)}P^{-\mu}_\nu(-z),
\label{730iga}
\eea
we obtain the asymptotics
\be
E_+(\mu,y)
\sim
\left\{
\begin{array}{ll}
e^{\mu y/b}
+R(\mu)
e^{-\mu y/b},
\qquad
 &y\to \infty,
\\[5pt]
T(\mu)
e^{\mu y/b},
\qquad
& y\to -\infty,
\end{array}
\right.
\label{734iga}
\ee
\be
E_-(\mu,y)
\sim
\left\{
\begin{array}{ll}
T(\mu)
e^{-\mu y/b},
\qquad
& y\to \infty,
\\[5pt]
e^{-\mu y/b}
+R(\mu)
e^{\mu y/b},
\qquad
 &y\to -\infty,
\end{array}
\right.
\label{735iga}
\ee
where
\bea
T(\mu) &=&
\frac{\Gamma(\mu+\nu+1)\Gamma(\mu-\nu)}{\Gamma(\mu+1)\Gamma(\mu)},
\\[5pt]
R(\mu) &=&
\frac{\sin(\pi\nu)}{\sin(\pi\mu)}
\frac{\Gamma(\mu+\nu+1)\Gamma(\mu-\nu)}{\Gamma(\mu+1)\Gamma(\mu)}.
\eea
This enables one to easily compute the
Wronskian of these solutions
\be
E_+(\mu,y)\partial_y E_-(\mu,y)
-\partial_y E_+(\mu,y) E_-(\mu,y)
=-\frac{2\mu}{b} T(\mu).
\ee

Therefore, for $\Real\mu>0$ the function $E_-(\mu,y)$
is bounded as $y\to \infty$ and the function $E_+(\mu,y)$
is bounded as $y\to -\infty$.
For imaginary $\mu=\pm ip$
these solutions behave like incoming and outgoing waves.

By using the relations (\ref{725iga}) it is easy to see that
these functions are related by
\bea
E_+(\mu,y)
&=&
R(\mu)E_+(-\mu,y)
+T(\mu)E_-(-\mu,y),
\\
E_-(\mu,y)
&=&
T(\mu)E_+(-\mu,y)
+R(\mu)E_-(-\mu,y),
\eea
or in the matrix form
\be
\begin{pmatrix}
E_+(\mu)\\
E_-(\mu)
\end{pmatrix}
=C(\mu)\begin{pmatrix}
E_+(-\mu)\\
E_-(-\mu)
\end{pmatrix},
\ee
where $C(\mu)$ is a symmetric matrix defined by
\be
C(\mu)=
\begin{pmatrix}
R(\mu) & T(\mu)
\\
T(\mu) & R(\mu)
\end{pmatrix}.
\ee
Therefore, the matrix $C(\mu)$ satisfies the
functional equation
\be
C(\mu)C(-\mu)=I.
\ee
In particular, this means
\bea
R(\mu)T(-\mu)+T(\mu)R(-\mu) &=& 0,
\label{744iga}\\
R(\mu)R(-\mu)+T(\mu)T(-\mu) &=& 1.
\label{745iga}
\eea
Obviously, it is symmetric and satisfies
\be
\overline{C(\mu)}=C(\bar\mu),
\qquad
C^T(\mu)=C(\mu);
\ee
therefore, for the imaginary $\mu$, with $\Real\mu=0$,
the matrix $C(\mu)$ is unitary,
\be
C(\mu)C^*(\mu)=I.
\ee
In particular, this means that
the matrix $C(\mu)$ does not have poles on the
imaginary axis.

We summarize the properties of the matrix $C(\mu)$.
Let $\nu$ be a positive real parameter.
Then:
\benum
\item
The matrix $C(\mu)$ is a meromorphic function of $\mu$.
\item
It is analytic for $\Real \mu >\nu$;
all poles are located in the half-plane
$\Real\mu\le \nu$.
\item
All poles are on the real axis.
\item
The poles on the interval $(0,\nu]$ are simple and have the form
\be
a_j=\nu -j, \qquad
j=0,1,\dots, [\nu].
\ee

\item
If $2\nu$ is not an integer, then there are two series of
simple poles in the
left half-plane $\Real\mu<0$,
\be
b_m=-\nu-1-m,
\qquad
m=0,1,2,\dots
\ee
and
\be
a_j=\nu -j, \qquad
j\ge [\nu]+1.
\ee

\item
If $2\nu$ is a positive integer then the poles
\be
a_j=\nu -j,
\qquad
[\nu]+1\le  j\le 2\nu,
\ee
are single
and the poles
\be
b_m=-\nu-1-m,
\qquad
m=0,1,2,\dots
\ee
are double.

\item
The matrix $C(\mu)$ satisfies the functional equation
\be
C(\mu)C(-\mu)=I
\ee
and is unitary on the imaginary axis, for $\Real\mu=0$.

\eenum

Therefore, the function $\tilde G_0(\lambda;y,y')$ is
given by
\bea
\tilde G_0(\lambda;y,y')
&=&
\frac{b}{2\mu T(\mu)}
E_+(\mu,y)E_-(\mu,y');
\eea
by using (\ref{726igax}) we see that
\be
\tilde G_0(\lambda;y,y')=\tilde G_0(\lambda;-y',-y).
\ee
The resolvent is
\be
G_0(\lambda;y,y')
=\theta(y'-y)\tilde G_0(\lambda;y,y')
+\theta(y-y')\tilde G_0(\lambda;y',y).
\ee
which is obviously symmetric and
has the correct asymptotics at infinity
(\ref{77iga}).
The asymptotics of the resolvent at infinity,
as $y,y'\to \infty$ or $y,y'\to-\infty$ is
\be
\tilde G_0(\lambda;y,y')
\sim
\frac{b}{2\mu}\left\{e^{\mu(y-y')/b}
+R(\mu)e^{-\mu|y+y'|/b}
\right\}.
\ee
The diagonal value of the resolvent is
\bea
G_0(\lambda;y,y)
&=&
\frac{b}{2\mu T(\mu)}
E_+(\mu,y)E_-(\mu,y).
\eea
which is an even function
\be
G_0(\lambda;y,y)=G_0(\lambda;-y,-y)
\ee
with the asymptotics as $|y|\to\infty$,
\be
G_0(\lambda;y,y)
\sim
\frac{b}{2\mu}\left\{1
+R(\mu)e^{-2\mu|y|/b}
\right\}.
\ee
Note that for $\Real\mu>0$ the resolvent diagonal approaches a constant at
infinity $|y|\to \infty$,
\be
G_0(\lambda;y,y)\sim \frac{b}{2\mu}
\ee
and, therefore, is not integrable.


The spectrum is determined by the singularities
of the resolvent.
The resolvent is a meromorphic function $\mu$.
It has a finite number of simple poles
on the positive real line at
\be
a_j=\nu-j,
\qquad
j=0,1,2\dots, N,
\ee
that is,
\be
\lambda_{0,j}=\frac{1}{b^2}\left[\nu^2-(\nu-j)^2\right]
=\frac{1}{b^2}j(2\nu-j),
\ee
where
$
N=\lceil \nu\rceil-1
$
is the greatest integer less than $\nu$.
The corresponding eigenfunctions are
\be
\psi_{0,j}(y)
=c_{0,j}P^{j-\nu}_\nu(z)
\ee
with some normalization constants.
These eigenfunctions decrease at infinity
exponentially as $y\to \pm\infty$,
\be
\psi_{0,j}(y)\sim \exp[-(\nu-j)|y|/b],
\ee
for any $j<\nu$.

By using the integral
\cite{erdelyi53}
\be
\int\limits_{-1}^1 \frac{dz}{1-z^2}
[P^{j-\nu}_\nu(z)]^2
=\frac{1}{2(\nu-j)}\frac{j!}{\Gamma(2\nu-j+1)}
\ee
we find
\be
c_{0,j}=\frac{1}{\sqrt{b}}
\left(2(\nu-j)\frac{\Gamma(2\nu-j+1)}{j!}\right)^{1/2}.
\ee

The corresponding eigenfunctions of the Laplacian
on the manifold $M$ are
\be
\Psi_{0,j}(y,\hat x)=\frac{c_{0,j}}{\sqrt{\vol(N)}}P^{j-\nu}_\nu(z).
\ee

It is obvious that, as a function of $\lambda$,
the resolvent has a branch cut singularity
along the positive real axis from $\nu^2$ to infinity,
which determines the continuous spectrum $[\nu^2/b^2,\infty)$.


Now, we obtain the heat kernel of the operator $D_0$
by the inverse Laplace transform (\ref{617iga}).
We deform the contour of integration to a contour
 in the complex plane of $\lambda$ which goes from
$-i\varepsilon+\infty$ around the positive real half-axis
to $i\varepsilon+i\infty$. Recall that the upper bank of the
branch cut corresponds to $\mu=-ip$ and the lower
bank of the cut corresponds to $\mu=ip$ with $p>0$.
This leads to the sum of the poles of the resolvent
and the integral over the branch cut of the jump of the resolvent
across the cut,
\bea
U_0(t;y,y') &=&
\sum_{j=0}^{N}\exp\left\{-[\nu^2-(\nu-j)^2]\frac{t}{b^2}\right\}
\psi_{0,j}(y)\psi_{0,j}(y')
\nn\\
&&
+\frac{1}{b}\int\limits_{0}^{\infty}\frac{dp}{2\pi}\;
\exp\left\{-(\nu^2+p^2)\frac{t}{b^2}\right\}
W(p;y,y'),
\label{565iga}
\eea
where
\bea
W(p;y,y')
&=& E_+(ip,y)E_+(-ip,y')
+E_-(ip,y)E_-(-ip,y').
\label{567iga}
\eea
This function has the properties
\be
W(p;y,y')=W(p;-y,-y')=W(-p,y,y'),
\ee
By using (\ref{734iga}), (\ref{735iga}),
and (\ref{744iga}), (\ref{745iga}), we get
as $y,y'\to \infty$ or $y,y'\to -\infty$,
\be
W(p;y,y')
\sim
2\cos[p(y-y')/b]
+R(ip)e^{-ip|y+y'|/b}+R(-ip)e^{ip |y+y'|/b}.
\ee
The diagonal value of this function
is a real even function
of both $y$ and $p$.
By using the definition of the functions
$E_\pm(\mu,y)$ and the properties of the gamma function
we compute
\bea
W(p;y,y)
&=&
\frac{\sinh^2(\pi p)}{\sinh[\pi(p-i\nu)]\sinh[\pi(p+i\nu)]}
\left\{
\Psi(p,y)+\Psi(p,-y)
\right\}
\label{570iga}
\eea
where
\be
\Psi(p,y)
=F\left(-\nu{},\nu{}+1;1+ip;\frac{1+z}{2}\right)
F\left(-\nu{},\nu{}+1;1-ip;\frac{1+z}{2}\right)
\ee
We will also need the asymptotics of this function as $p\to\infty$.
By using the properties of the hypergeometric function we get
\be
\Psi(p,y)= 1-\frac{\nu(\nu+1)}{p^2}
\left\{1
-(\nu-1)(\nu+2)\frac{(1+z)}{2}
\right\}\frac{(1+z)}{2}
+O(p^{-3})
\ee
Therefore, as $p\to \infty$,
\be
W(p;y,y)
\sim
 2-\frac{\nu(\nu+1)}{p^2}
\left\{1
-\frac{(\nu-1)(\nu+2)}{2}
(1+z^2)
\right\}
+O(p^{-3})
\ee

Note that as $p \to 0$,
\be
W(p;y,y) \sim \frac{\pi^2p^2}{\sin^2(\pi\nu)}
\left\{
\Psi(0,y)+\Psi(0,-y)
\right\},
\ee
and as $|y|\to \infty$,
\be
W(p;y,y)\sim
2+R(ip)e^{-2ip |y|/b}
+R(-ip)e^{2ip |y|/b}.
\ee
where
\be
R(ip) =
\sin(\pi\nu)\frac{p}{\sinh(\pi p)}
\frac{\Gamma(ip+\nu+1)\Gamma(ip-\nu)}{\left[\Gamma(ip+1)\right]^2
}.
\ee
Notice that as $p\to 0$
\be
R(ip)= -1-cip+O(p^2)
\ee
where
\be
c=\left[\psi(\nu+1)+\psi(-\nu)
-2\psi(1)\right],
\ee
where $\psi(z)=\Gamma'(z)/\Gamma(z)$ is the logarithmic derivative
of the gamma function.

Next, by using the asymptotics of the gamma function
\be
\Gamma(z) \sim (2\pi)^{1/2}z^{-1/2}\exp\left\{
z\log z -z
\right\}
\ee
we get
\be
\frac{\Gamma(z+a)}{\Gamma(z)} \sim z^a
\ee
and, therefore,
as $p\to \infty$
\be
R(ip) \sim
-i\frac{\sin(\pi\nu)}{\sinh(\pi p)}.
\ee



The operator $D_0,$ in the compact case, $\Sigma=S^1$,
has a discrete spectrum with a well defined heat trace
\be
\Tr \exp(-tD_0)
=\sum_{j=1}^\infty d_{0,j}e^{-t\lambda_{0,j}}.
\label{411igaxx}
\ee
However, in the non-compact case, $\Sigma=\RR$, contrary
to the operators $D_k$ with $k\ge 1$, the operator
has both discrete and continuous spectrum. In this case
the heat trace $\Tr\exp(-tD_0)$ does not exist;
it needs to be regularized.
If there are finitely many discrete simple eigenvalues
$\lambda_{0,j}$, $j=1,\dots, N$, in the interval
$[0, m^2]$ with some positive constant $m$,
and the rest of the spectrum is continuous,
then the regularized heat trace is
\be
\Tr_{\rm reg}\exp(-tD_0) =
\sum_{j=1}^N e^{-\lambda_{0,j}t}
+\int\limits_{0}^{\infty} \frac{d{}p}{2\pi}\;
\exp\left\{-(m^2+{}p^2)t\right\}
V_{\rm reg}({}p),
\ee
where
\be
V_{\rm reg}({}p)=
-2i{}p
\int\limits_\RR dy
\;{\rm Jump}\;G^{\rm reg}_0\left(m^2+p^2;y,y\right)
\ee
where
\be
{\rm Jump}\;G^{\rm reg}_0\left(m^2+p^2;y,y\right)
=G^{\rm reg}_0\left(m^2+p^2+i\varepsilon;y,y\right)
-G^{\rm reg}_0\left(m^2+p^2-i\varepsilon;y,y\right)
\ee
is the jump of the regularized resolvent
\be
G_0^{\rm reg}(\lambda;y,y)=
G_0(\lambda;y,y)
-\lim_{|y|\to\infty}G_0(\lambda;y,y)
\ee
across the cut.

In particular, in the case of two cusps manifold (\ref{212iga})
with the heat kernel given by (\ref{565iga}),
the
regularized heat trace of the operator $D_0$
is
\bea
\Tr_{\rm reg}\exp(-tD_0)
&=&
\sum_{j=0}^{N}\exp\left\{-\left[\nu^2-(\nu-j)^2\right]\frac{t}{b^2}\right\}
\label{783iga}\\
&&
+
\int\limits_{0}^{\infty}\frac{d{}p}{2\pi}\;
\exp\left[-(\nu^2+{}p^2)\frac{t}{b^2}\right]
\int\limits_\RR dy\;
\frac{1}{b}\left[W({}p;y,y)-2\right].
\nn
\eea
where
$
W({}p;y,y)
$
is defined in (\ref{570iga}).




The asymptotics of the heat kernel diagonal
$U_0(t;y,y)$ of the operator $D_0$ has the same form
(\ref{521iga}),
\be
U_0(t;y,y)\sim
(4\pi )^{-1/2}
\exp\left[-tQ_0(y)\right]
\sum_{j=0}^\infty \frac{t^{j-1/2}}{j!}b_{0,j}(y),
\label{91igaxt}
\ee
where the coefficients $b_{0,j}$ are differential polynomials
of the potential $Q_0$ of degree $j$.
The potential $Q_0$ is defined by (\ref{37iga}).
We suppose that the function $\omega$
grows at infinity like $c|y|$ so that the potential
$Q_0$ approaches a constant $m^2$ at infinity,
that is, we decompose the potential $Q_0$
by separating its limit at infinity,
\be
Q_0(y)=m^2+u(y),
\ee
where $u(y)$ is a smooth function decreasing
exponentially at infinity
together with its derivatives.
Since the term $\exp[-tQ_0(y)]$ does not provide
the convergence of the integral over $y$, it does not make sense
to keep it in the exponential, but rather to use the
asymptotic expansion (\ref{56iga})
\be
U_0(t;y,y)\sim
(4\pi )^{-1/2}\exp\left(-tm^2\right)
\sum_{j=0}^\infty \frac{t^{j-1/2}}{j!}c_{j}(y),
\label{91igaxy}
\ee
where the coefficients $c_j$ are
differential polynomials of the function $u$ of degree $j$
defined by (\ref{59iga}).
The coefficients $c_{j}(y)$ for $j\ge 1$
(except for the first one $c_{0}=1$)
are exponentially small at infinity and, therefore,
are integrable
except for the first one $c_0=1$.
Therefore, the global heat kernel coefficients
\be
C_k(\RR)=\int\limits_{\RR} dy\; c_k(y),
\qquad k\ge 1,
\ee
are well defined; they have the form
\bea
C_1(\RR) &=& -\int\limits_{\RR} dy\; u,
\\
C_2(\RR) &=& \int\limits_{\RR} dy\;
u^2,
\\
C_k(\RR) &=&
\int\limits_{\RR} dy\;
\left\{
\frac{k!(k-1)!}{(2k-2)!}u\partial_y^{k-2}u
+\cdots
+(-1)^ku^k
\right\},
\qquad
k\ge 2.
\eea


In particular,
in the case of cusps, with the function $\omega$
defined by (\ref{212iga}), the potential $Q_0$ is
given by (\ref{525iga})
and approaches the constant $\nu^2/b^2$ at infinity,
\be
Q_0(y) =
\frac{\nu^2}{b^2}
+u(y),
\ee
where
\be
u(y)=
-\frac{\nu(\nu+1)}{b^2\cosh^2(y/b)}.
\label{525igax}
\ee
In this case these coefficients can be computed
exactly
\bea
C_1(\RR) &=& 2\nu(\nu+1)\frac{1}{b},
\\
C_2(\RR) &=&
\frac{4}{3}\nu^2(\nu+1)^2\frac{1}{b^3}.
\eea
It is easy to see that
\be
C_k(\RR)=b^{1-2k}P_k(\nu),
\ee
where $P_k(\nu)$ are polynomials in $\nu$ of degree $2k$.

The heat kernel diagonal $U_0(t;y,y)$ is not integrable
and the heat trace $\Tr\exp(-tD_0)$ does not exist.
Therefore, it needs to be regularized by removing
the first term $c_{0}$.
The regularized heat trace of the operator $D_0$
defined by
\bea
\Tr_{\rm reg} \exp(-tD_0)
&=&
\int\limits_{\RR}dy\;
\left\{U_0(t;y,y)
-(4\pi t)^{-1/2}e^{- tm^2}
\right\}
\label{732iga}
\eea
has the asymptotic expansion
\bea
\Tr_{\rm reg} \exp(-tD_0)
&\sim &
(4\pi)^{-1/2}\exp(-tm^2)
\sum_{k=1}^\infty \frac{t^{k-1/2}}{k!}C_k(\RR).
\label{732igam}
\eea
By comparing this equation in the case of cusps with
$m^2=\nu^2/b^2$
with equation (\ref{783iga}) we obtain a nontrivial
relation
\bea
&&\sum_{j=0}^{N}\exp\left\{(\nu-j)^2t\right\}
+\int\limits_{0}^{\infty}\frac{d{}p}{2\pi}\;
\exp\left(-{}p^2t\right)
\int\limits_\RR dy\;\frac{1}{b}\left[
W({}p;y,y)-2\right]
\nn\\
&&\sim(4\pi )^{-1/2}
\sum_{k=1}^\infty \frac{t^{k-1/2}}{k!}P_k(\nu).
\label{783igaz}
\eea


\section{Heat Trace $\Tr\exp(t\Delta_M)$}
\setcounter{equation}{0}

It is well known that
in the case of a compact manifold without boundary
 the heat trace $\Tr\exp(t\Delta_M)$ of the Laplacian
$\Delta_M$ has the  asymptotic expansion (\ref{11iga}),
\cite{gilkey95,avramidi00},
 as $t\to 0$
\be
\Tr\exp(t\Delta_M)
\sim (4\pi)^{-n/2}\sum_{k=0}^\infty
\frac{t^{k-n/2}}{k!} A_k(M),
\ee
where
\bea
A_k(M) &=& \int\limits_M d\vol_M a_k
\nn\\
&=&
\int\limits_\RR dy
\exp\left[-2\alpha\omega(y)\right]
\int\limits_N d\vol_N(\hat x) a_k(y,\hat x),
\eea
and $a_k$  are differential polynomials in the curvature
of the manifold $M$ of degree $k$, that is,
they are polynomials in the curvature
and its covariant derivatives of dimension $L^{-2k}$,
with $L$ the unit length scale.
That is, they are polynomials in the function $e^{2\omega}$,
the derivatives of the function $\omega$ and the curvature and its
derivatives of the manifold $N$, that is,
\bea
A_k(M) &=&
\int\limits_{\Sigma}dy
\sum_{l=0}^k \exp\left[-2(\alpha-l)\omega(y)\right]
a_{k,l}(y),
\eea
where $a_{k,l}(y)$ are differential polynomials of the function $\omega$
of degree $k$.

In the case of the
compact warped product manifold $M=S^1\times_f N$
all these global coefficients are well defined;
they have the following general
form
\be
A_k(M)
=\sum_{j=0}^k C_{k,k-j}(S^1)
B_{k,j}(N),
\ee
where $B_{k,j}(N)$ are integrals over the manifold $N$
of scalar invariants of the curvature $F_{ijkl}$
of the manifold $N$
and its covariant derivatives
of dimension $L^{-2j}$ and
$C_{k,k-j}(S^1)$ are integrals over $S^1$ of
 differential polynomial of the function $\omega$
and $e^{-\omega}$ of dimension $L^{-2(k-j)}$.
In particular, by using the well known
asymptotics \cite{avramidi00} and (\ref{226iga}),
we get
\bea
A_0(M) &=& \vol(M)
\nn\\
&=&C_{0,0}(S^1)\vol(N),
\\
A_1(M) &=&
\frac{1}{6}\int\limits_M d\vol_MR
\nn\\
&=&C_{1,2}(S^1)\vol(N)
+C_{1,0}(S^1)
\int\limits_N d\vol_N F,
\eea
where $R$ is the scalar curvature of the manifold $M$,
$F$ is the scalar curvature of the manifold $N$ and
\bea
C_{0,0}(S^1) &=& \int\limits_{S^1}dy\;e^{-2\alpha\omega},
\\
C_{1,2}(S^1) &=&
\int\limits_{S^1}dy\;e^{-2\alpha\omega}
\frac{\alpha}{3}\left\{
2\omega''
-(2\alpha+1)\omega'^2
\right\},
\\
C_{1,0}(S^1) &=&\frac{1}{6}
\int\limits_{S^1}dy e^{-2(\alpha-1)\omega}.
\eea


In the case of a noncompact warped product manifold
$M=\RR\times_f N$ of finite volume
the heat trace  of the
Laplacian $\Tr\exp(t\Delta_M)$ on the manifold $M$
needs to be regularized.
By using eq. (\ref{47iga}) we have
\be
\Tr_{\rm reg}\exp(t\Delta_M)=\Tr_{\rm reg}\exp(-tL),
\ee
where $L$ is the operator defined by (\ref{35iga}).
By separating the zero mode and using eq. (\ref{329iga})
we get
\be
\Tr_{\rm reg}\exp(t\Delta_M) = \Tr_{\rm reg} \exp(-tD_0)
+\Tr\exp(-t\tilde L),
\label{82iga}
\ee
where the operator $D_0$ is defined by (\ref{39iga}),
the operator $\tilde L$ is defined by (\ref{332iga}),
and the regularized heat trace
$\Tr_{\rm reg}\exp(-tD_0)$ of the operator $D_0$
is defined
by (\ref{732iga})
with the asymptotic expansion (\ref{732igam}).
%



The heat trace $\Tr\exp(-\tilde L)$ is given by
\bea
\Tr\exp(-t\tilde L) &=&
\Tr\exp\left[-t\left(D_0-e^{2\omega}\tilde \Delta_N\right)\right]
\eea
We use the eq. (\ref{339iga}) to obtain
the
heat trace
\be
\Tr\exp(-t\tilde L)
=\int\limits_\RR dy
\int\limits_0^\infty d\tau\;
\Tr\exp(\tau\tilde\Delta_N)
\int\limits_{c-i\infty}^{c+i\infty}
\frac{d\lambda}{2\pi i}\;
e^{\tau\lambda}
\Phi(t,\lambda;y,y).
\ee
The heat kernel diagonal $\Phi(t,\lambda;y,y)$
has the asymptotic expansion
as $t\to 0$
\be
\Phi(t;\lambda,y,y)
\sim
(4\pi )^{-1/2}
\exp\left(
-t\lambda e^{2\omega}\right)
\sum_{k=0}^\infty t^{k-1/2}
\sum_{j=0}^k
\left(\lambda e^{2\omega}\right)^j
\Omega_{k,j},
\ee
where $\Omega_{k,j}$ are differential polynomials
of $\omega$.


Now, by using the integrals
\be
\int\limits_{c-i\infty}^{c+i\infty}
\frac{d\lambda}{2\pi i}\;
\exp\left[\lambda\left(\tau-te^{2\omega}
\right)\right]
\lambda^j
=\partial_\tau^j\delta(\tau-te^{2\omega})
\ee
we obtain
\be
\Tr\exp(-t\tilde L)
\sim(4\pi )^{-1/2}
\sum_{k=0}^\infty t^{k-1/2}
\sum_{j=0}^k X_{k,j}(t),
\ee
where
\be
X_{k,j}(t)=(-\partial_t)^j
\int\limits_\RR dy\; \Omega_{k,j}
\Tr\exp(te^{2\omega}\tilde\Delta_N).
\label{339igaxy}
\ee
Finally, we express the heat trace in terms of the zeta function and obtain
\be
X_{k,j}(t)=
\int\limits_{\sigma-i\infty}^{\sigma+i\infty}
\frac{ds}{2\pi i}t^{-s-j}
\frac{\Gamma(s-\alpha)\Gamma(s+j)}{\Gamma(s)}
Z_N(s)
\int\limits_\RR dy\; e^{-2s\omega} \Omega_{k,j}.
\label{339igaxyz}
\ee
where $\sigma>\alpha$.


In the case of a noncompact warped product manifold
$M=\RR\times_f N$ of finite volume the local coefficients
$a_k$ are of order $R^k$ (with $R$ being the
curvature) and, therefore, behave like
$\exp(2k\omega)$ at infinity, as $|y|\to \infty$.
However, since the volume element $d\vol_M$ behaves like
$\exp(-2\alpha\omega)$,
the coefficients
$A_k(M)$ with $k<\alpha$ are, in fact, well defined,
that is, the first coefficients of the asymptotic expansion
are the same in the compact and in the noncompact case
\be
\Tr\exp(t\Delta_M)
= (4\pi)^{-n/2}\sum_{k=0}^{\varkappa}
\frac{t^{k-n/2}}{k!} A_k(M)+\Theta_{M,+}(t),
\ee
where $\varkappa=[n/2]-1$ and
$\Theta_{M,+}(t)$ is a function of order
$o\left(t^{\varkappa-n/2}\right)$,
that is, of order $o\left(t^{-1}\right)$ if $n$ is even,
and of order $o\left(t^{-3/2}\right)$
if $n$ is odd.
The leading asymptotics has the form
\be
\Theta_{M,+}(t)
=   t^{-1/2}\left\{S_1(M)\log t
+ S_2(M) \right\}
+O(t^{1/2});
\ee
due to the non-compactness of the manifold $M$
the coefficients $S_1(M)$ and $S_2(M)$
of this expansion are non-local.

In the case of cusps, one can compute the
asymptotics explicitly.
By using (\ref{418igax}) we obtain
\bea
S_1(M) &=&
-b(4\pi)^{-1/2}
\frac{\alpha}{\nu}\zeta_N(0),
\\
S_2(M) &=&
b(4\pi)^{-1/2}
\left\{
\frac{\alpha}{\nu}\zeta'_N(0)
+\left(\frac{\alpha}{\nu}\gamma+2\right)
\zeta_N(0)\right\}.
\label{418igaxy}
\eea




\section*{Acknowledgement}

This work was initiated during a visit of the author at the University of Bonn.
The author would like to thank the Mathematical Institute of the University
of Bonn, in particular, Werner M\"uller, for hospitality.

\ed

\end{document}